\documentclass[aps,pre,twocolumn,superscriptaddress,superscriptreference]{revtex4-2}
\usepackage{amsmath,bbold,bm,amssymb,scalerel,mathtools}
\usepackage{graphicx}
\usepackage{color}
\usepackage{enumitem}
\usepackage{algorithm,algpseudocode}
\usepackage{multirow}
\usepackage{colortbl,booktabs}
\usepackage{placeins}
\usepackage[usenames,dvipsnames]{xcolor}
\usepackage{tikz}
\usepackage{lipsum} 
\usepackage[colorlinks, linkcolor=blue!50!black, urlcolor=blue!50!black, citecolor=blue!50!black]{hyperref}
\usepackage{cleveref}
\crefname{equation}{Eq.}{Eqs.} 
\crefname{section}{Sec.}{Secs.} 
\crefname{figure}{Fig.}{Figs.}
\crefname{table}{Tab.}{Tabs.} 
\crefname{appendix}{App.}{App.}
\tikzset{%
	every neuron/.style={
		circle,
		draw,
		minimum size=0.5cm
	},
	neuron missing/.style={
		draw=none, 
		scale=1.5,
		text height=0.333cm,
		execute at begin node=\color{black}$\vdots$
	},
}

\newcommand\equalhat{%
	\let\savearraystretch\arraystretch
	\renewcommand\arraystretch{0.3}
	\begin{array}{c}
		\stretchto{
			\scalerel*[\widthof{=}]{\wedge}
			{\rule{1ex}{3ex}}%
		}{0.5ex}\\ 
		=%
	\end{array}
	\let\arraystretch\savearraystretch
}

\usepackage[normalem]{ulem}

\begin{document}

\title{Dynamic heterogeneity at the experimental glass transition predicted by transferable machine learning}

\author{Gerhard Jung}

\affiliation{Laboratoire Charles Coulomb (L2C), Université de Montpellier, CNRS, 34095 Montpellier, France}

\author{Giulio Biroli}

\affiliation{Laboratoire de Physique de l’Ecole Normale Supérieure, ENS, Université PSL, CNRS, Sorbonne Université, Université de Paris, F-75005 Paris, France}

\author{Ludovic Berthier}

\affiliation{Laboratoire Charles Coulomb (L2C), Université de Montpellier, CNRS, 34095 Montpellier, France}

\affiliation{Gulliver, UMR CNRS 7083, ESPCI Paris, PSL Research University, 75005 Paris, France}

\date{\today}

\begin{abstract}
We develop a transferable machine learning model which predicts structural relaxation from amorphous supercooled liquid structures. The trained networks are able to predict dynamic heterogeneity across a broad range of temperatures and time scales with excellent accuracy and transferability. We use the network transferability to predict dynamic heterogeneity down to the experimental glass transition temperature, $T_g$, where structural relaxation cannot be analyzed using molecular dynamics simulations. The results indicate that the strength, the geometry and the characteristic length scale of the dynamic heterogeneity evolve much more slowly near $T_g$ compared to their evolution at higher temperatures. Our results show that machine learning techniques can provide physical insights on the nature of the glass transition that cannot be gained using conventional simulation techniques.    
\end{abstract}

\maketitle

\section{Introduction}

Dense liquids display a drastic slowing down of structural relaxation when approaching the experimental glass transition temperature~\cite{donth2001glass,debenedetti2001supercooled}. The glass transition is characterized by several important properties, such as a very homogeneous amorphous structure but a strongly heterogeneous relaxation dynamics, leading to the spatial coexistence of frozen and active regions~\cite{bookDH}. Understanding the connection between the microstructure and dynamic heterogeneity is an important field of research~\cite{widmer2008irreversible,Royall2015,tanaka2019revealing,PhysRevMaterials.4.113609}.

Over the years, several structural order parameters have been proposed which show some degree of correlation with the local relaxation dynamics, including density~\cite{unsupervisedCoslovich}, potential energy~\cite{PhysRevLett.91.235501,Harrowell2006}, geometry of Voronoi cells~\cite{PhysRevLett.116.088001}, soft modes~\cite{widmer2008irreversible}, locally-favored structures~\cite{Malins2013,tong2018revealing,hocky2014correlation,Royall2015}, and more~\cite{lerbinger2021relevance,tanaka2019revealing,PhysRevMaterials.4.113609,PhysRevLett.127.088002}. Recently, the application of machine learning (ML) techniques to automatically construct suitable structural order parameters has significantly advanced this line of research. The range of methodologies includes unsupervised learning to automatically detect structural heterogeneities~\cite{unsupervisedFilion,unsupervisedCoslovich,CNN2022,PhysRevE.106.025308} and supervised learning using linear regression~\cite{Filion2021,Filion2022,alkemade2023improving}, support vector machines~\cite{PhysRevLett.114.108001,schoenholz2016structural,Zaccone2021}, multilayer perceptrons (MLP)~\cite{jung2023_PRL} and graph neural networks (GNN)~\cite{bapst2020unveiling,pezzicoli2022se,GNNrelative2022,jiang2022geometry}. The performance of these techniques significantly surpasses traditional approaches based on hand-made order parameters, and allows to infer the microscopic structural relaxation from structural properties with high accuracy, including aspects of dynamic heterogeneity~\cite{jung2023_PRL,pezzicoli2022se}. 

Thanks to this progress, ML approaches lead to new physical results. Applications of trained neural networks have used scalability in system size to extract new results on dynamic length scales and the geometry of rearranging domains~\cite{jung2023_PRL}, transferability to other state points to analyze structural differences between strong and fragile glass formers~\cite{10.1063/5.0099071}. Trained models were also used to construct effective glass models~\cite{PhysRevResearch.4.043026}. While the performance of ML approaches is remarkable, one of the main drawbacks of the supervised learning techniques is that they need to be trained separately for each state point, thus requiring that training sets already exist at each time and temperature. Transferability to lower temperature has been analyzed in one of the first ML applications~\cite{bapst2020unveiling,pezzicoli2022se}. It was shown that for GNNs some correlation between structure and dynamics persists when applying the trained networks to different temperatures~\cite{bapst2020unveiling,pezzicoli2022se}. However, a full analysis of transferability is lacking. The aim of this article is to fill this gap, and present an efficient method to predict physical properties outside of the training regime, possibly including physical regimes that cannot easily be accessed using conventional numerical techniques.  

We develop a transferable ML framework, which is able to learn and predict time-dependent dynamical properties when given amorphous structures of deeply supercooled liquids. Different from most previously proposed ML techniques, the network is trained using data extracted from very different temperatures and time scales, and is thus able to maintain both an excellent performance over the whole range of data provided. We show that the network can be transferred to predict relaxation beyond the range of temperatures provided during training. As an important application, we create equilibrium structures down to the experimental glass transition temperature, $T_g,$ using the swap Monte-Carlo algorithm~\cite{swap:ninarello2017,swap:Berthier2019} and apply the transferable network to investigate dynamic susceptibilities and lengths scales at these very low temperatures, which correspond to time scales that are not accessible by molecular dynamics (MD) simulations.

The manuscript is organized as follows. First, the transferable ML methodology is described in \cref{sec:methodology}. The performance of the trained network is analyzed in \cref{sec:performance}, both for state points within the range of training data and beyond to study transferability. In \cref{sec:glass_transition} we use the transferability to obtain new results for structural relaxation at the experimental glass transition temperature. In \cref{sec:length_scales}, we analyze in detail the dynamic correlation lengths of our system. To better understand the trained network, we study some of its intrinsic properties, such as inherent length scales and properties of the bottleneck layer in \cref{sec:network}.  We discuss the results and conclude in \cref{sec:discussions}. 

\section{Machine Learning Methodology}

\label{sec:methodology}

\begin{figure}
	\includegraphics[width=\linewidth]{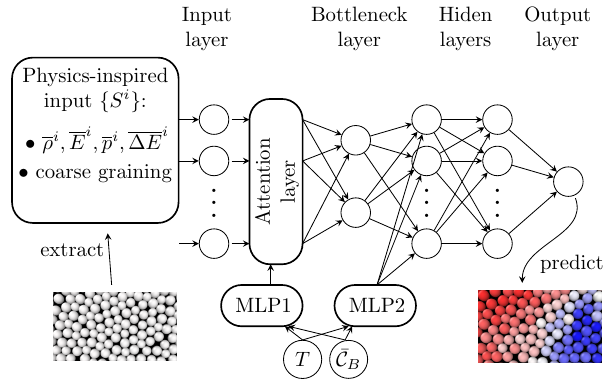}
	\caption{Geometry of the transferable machine learning model, which we name `tGlassMLP'. The various layers and their respective role are described in \cref{sec:methodology}.}
	\label{fig:geometry}
\end{figure}

The transferable ML technique presented in this work is a generalization of the GlassMLP network introduced in Ref.~\cite{jung2023_PRL} which we modify and improve to enable transferability. We will refer to this model as `transferable GlassMLP' and use the acronym `tGlassMLP' for the model. The geometry of tGlassMLP is sketched in \cref{fig:geometry}. It is comprised of an input layer, an attention layer, and a dense multi-layer perceptron (MLP) network including a bottleneck layer. We now describe these individual parts, define the glass-forming system under study and the dynamical observables used to train the network.

\subsection{Physical system and dynamical observables}

\label{sec:observables}

We study the same system and dynamical observables used in Ref.~\cite{jung2023_PRL}, but we briefly recapitulate the most important details and definitions to make the paper self-contained.

We investigate a two-dimensional ternary glass-former, which generalises the Kob-Andersen binary binary mixture (KA2D)~\cite{PhysRevE.51.4626}. To enable fast equilibration using the swap Monte-Carlo algorithm~\cite{berthier2016equilibrium,swap:ninarello2017,swap:Berthier2019} we include a third particle type with an intermediate size~\cite{Berthier2020}. We use reduced units, which are defined in terms of length $\sigma=1$ (corresponding to the size of the large particles), mass $m=1$ (mass of the particles) and energy $\epsilon=1$ (Lennard-Jones energy scale of the interactions between large particles). In these units, the standard system size is $L_S=32.896$, with periodic boundary conditions, in which we create amorphous packings of $N=1290$ particles ($N_1=600$, $N_2=330$, $N_3=360$), if not otherwise stated. Using swap Monte Carlo, we create equilibrium configurations in a temperature range between $T=0.4$, which is slightly below the onset temperature ($T_\text{on} \approx 0.5$), and the estimated experimental glass transition temperature $T_g=0.15$ (see SM in Ref.~\cite{jung2023_PRL} for details). Equilibrium averages over these configurations are denoted as $\langle \ldots \rangle.$ No sign of crystallization was observed even at $T_g.$ Starting from the equilibrated configurations we perform molecular dynamics (MD) simulations for five different temperatures, $T=0.4$, $T=0.3$, $T=0.25$, $T=0.23$ and $T=0.21,$ to investigate the physical relaxation dynamics. We analyze the isoconfigurational average~\cite{isoconf1,isoconf2}, which means that we perform $N_R=20$ different simulations starting from each structure (denoted as replicas) which are created by randomly drawing initial velocities from the Maxwell distribution. The isoconfigurational average $\langle \ldots \rangle_\text{iso}$ for a given initial structure denotes the average over these $N_R$ replicas.

The labels used to train tGlassMLP are extracted from the MD trajectories using the bond-breaking correlation function (BB). The isoconfigurational average of BB, which we refer to as `propensity' in the following~\cite{isoconf1,isoconf2}, is defined as $\mathcal{C}^i_B(t) = \langle n^i_t/n^i_0 \rangle_\text{iso}$, where $n^i_0$ denotes the initial number of neighbors of particle $i$, and $n^i_t$ the number of these neighbors which remain neighbors of $i$ after a time $t$~\cite{guiselin2022microscopic}. Therefore, $\mathcal{C}^i_B(t) \in [0,1]$, where $\mathcal{C}^i_B(t)=1$ denotes arrested particles which did not loose a single neighbor in any of the replicas (visualized as blue in all snapshots) and $\mathcal{C}^i_B(t)\ll1$ denotes very active particles (visualized as red). From the averaged propensity $\bar{\mathcal{C}}_B(t) = \frac{1}{N_1} \sum_{i =1 }^{N_1} \mathcal{C}^i_B(t)$, we extract a bond-breaking structural relaxation time, $\tau_\alpha^\text{BB}$, which is defined as $\langle \bar{\mathcal{C}}_B(t=\tau_\alpha^\text{BB}) \rangle = 0.5$. More details are documented in the SM of Ref.~\cite{jung2023_PRL}. We mostly report results for particles of type 1 but verified that all findings are independent of the particle type. We calculate BB from MD simulations for each temperature introduced above at various times, yielding many sets of labels at different state points for the supervised learning procedure of tGlassMLP.

To describe the state point of each set of labels, we use the temperature $T$ and the averaged propensity $\langle \bar{\mathcal{C}}_B(t) \rangle$. In particular, using $\langle \bar{\mathcal{C}}_B(t) \rangle$ instead of the time $t$ itself is a crucial choice due to a near perfect time-temperature superposition, i.e. perfect collapse of the function $\langle \bar{\mathcal{C}}_B(t/\tau_\alpha^\text{BB}) \rangle$ measured at different temperatures. Interchanging time for the value of $\langle \bar{\mathcal{C}}_B(t) \rangle$ simplifies the training procedure significantly since  $\langle \bar{\mathcal{C}}_B(t) \rangle \in [0,1]$, while $t$ grows exponentially with decreasing temperature. To translate $\langle \bar{\mathcal{C}}_B(t) \rangle$ back into a time, we assume that time-temperature superposition continues to hold at any temperature, combined with the extrapolated value for $\tau_\alpha^\text{BB}.$ More details of this procedure are provided in \cref{ap:BB}. While this assumption might affect slightly the time-dependence of the ML predictions presented in Secs.~\ref{sec:glass_transition}-\ref{sec:network}, they do not change the quality of the predictions themselves. Thus, the above encoding of time does not affect our analysis of transferability.

\subsection{Physics-inspired structural input}

The structural input is the same as for GlassMLP~\cite{jung2023_PRL}. In particular, we use $M_S$ physics-inspired and coarse-grained input features for the description of the local structure of each particle $i$. The descriptors are based on $K=4$ different observables:
\begin{enumerate}
	\item The coarse-grained local density: 
	\begin{equation}
	\overline{\rho}^i_{L,\beta} = \sum_{j\in N^i_\beta} e^{-R_{ij}/L},
	\end{equation}
	which sums over the $N_\beta^i$ particles of type $\beta$ within distance $R_{ij} = |\bm{R}_i - \bm{R}_j| < 20$ of particle $i$;
	\item The coarse-grained potential energy:
	\begin{equation}
	\overline{E}^i_{L,\beta} = \sum_{j\in N_\beta^i} E^j e^{-R_{ij}/L} / \bar{\rho}^i_{L,\beta},
	\end{equation}
	extracted from the pair interaction potential $E^i = \sum_{j\neq i} V(R_{ij})/2$;
	\item The coarse-grained Voronoi perimeter:
	\begin{equation}
	\overline{p}^i_{L,\beta} = \sum_{j\in N_\beta^i} p^j e^{-R_{ij}/L} / \overline{\rho}^i_{L,\beta} ,
	\end{equation}
	based on the perimeter $p^i$ of the Voronoi cell around particle $i,$ extracted using the software Voro++\cite{voro++};
	\item The local variance of potential energy:
	\begin{equation}
	\overline{\Delta E}^i_{L,\beta} = \sum_{j\in N_\beta^i} (E^j - \overline{E}_{L,\beta}^i)^2 e^{-R_{ij}/L} / \overline{\rho}^i_{L,\beta} .
	\end{equation}
\end{enumerate}   

Particle positions are evaluated in inherent structures $\bm{R}_i$. Slightly differently from Ref.~\cite{jung2023_PRL}, we use only $M_\text{CG}=11$ different values of the coarse-graining lengths $L$, which are non-uniformly distributed: $L=\{0.0, 1.0, 1.5, 2.0,2.5,3.0, 4.0, 5.0, 6.0, 7.0, 9.0\}$. The four descriptors are separately coarse-grained by iterating over each of the $M_\text{type}=3$ particle types. We additionally calculate the coarse-grained average by running over all particles independently of their type.

Overall, we start with a set of $M_S = K M_\text{CG} (M_\text{type}+1) = 176$ descriptors. To enable more efficient training, each descriptor is shifted and rescaled to have zero mean and unit variance.

\subsection{Attention layer}\label{sec:attention}

The first real difference between tGlassMLP and its GlassMLP ancestor is the introduction of an attention layer between the input and bottleneck layers (see \cref{fig:geometry}). The concept of attention and transformers have increased significantly the performance of many ML models in computer science~\cite{vaswani2017attention}, and it has already been used in glass physics for the development of an improved GNN~\cite{jiang2022geometry}. The purpose of the attention layer in tGlassMLP is to learn a state-dependent weight $w_j$ which is assigned to each structural descriptor $j$. In this way, the dependence of the dynamical descriptor on the considered state point is efficiently encoded in the network.

If we denote the $M_S$ values of the physics-inspired descriptors by $\{ S^1, ..., S^{M_S} \}$, then the attention layer can be written as,
\begin{align}
A^k &= S^k f_1^k + \text{MLP1}^1_\text{out} f_2^k + \text{MLP1}^2_\text{out} f_3^k + f_4^k\\
w_k &= \text{softmax}(\{A^1,..., A^{M_S}\})_k\\
S_\text{out}^k &= S^k w_k.
\end{align}
Here, $f_n^k$ denote the learnable parameters ($4 \cdot M_S = 688$ in total) and $\text{MLP1}^n_\text{out}$ is the two-dimensional output of a small MLP (denoted as MLP1). The softmax function ensures that the weights are normalized. The output of the attention layer still has dimension $M_S$, where each input descriptor, $S^k$ is multiplied by its specific weight $w_k$. The attention layer is able to reweight the input before encoding it in the bottleneck layer. It can thus learn, for example, that the relative importance of different structural indicators depends on time and temperature. We will explicitly analyze these weights $w_k$, obtained after training tGlassMLP, in \cref{sec:network} to extract meaningful physical information from interpreting the network itself.

\subsection{Dense MLP with bottleneck}

Similar to GlassMLP, after the attention layer, the high-dimensional input is encoded into a two-dimensional bottleneck layer, to avoid having a huge amount of free parameters in the subsequent hidden layers, which would lead to overfitting. We visualize and interpret the bottleneck layer of trained tGlassMLP networks in \cref{sec:network}.

After the bottleneck layer, the state point ($T,\langle \bar{\mathcal{C}}_B(t) \rangle$) is again explicitly inserted into the network using a second small MLP (denoted as MLP2). The bottleneck layer concatenated with the output of MLP2 will then be further processed in two hidden layers to yield the final output (see \cref{fig:geometry}). 

\subsection{Two-step training of tGlassMLP}

The total number of free parameters of tGlassMLP is slightly above 1000. This number is several orders of magnitude less than for the GNNs proposed in Refs.~\cite{bapst2020unveiling,pezzicoli2022se,GNNrelative2022,jiang2022geometry}. Importantly also, this number is not much larger than in the original version of GlassMLP where about $650$ fitting parameters were used~\cite{jung2023_PRL}. To be as efficient as GlassMLP, which was separately trained at each state point, the tGlassMLP network is therefore inherently forced to learn universal aspects in the structural relaxation across time scales and temperatures. This appears instrumental to construct a model with good transferability. 

To train the network, we use a supervised ML procedure, in which the output of tGlassMLP, $\mathcal{X}_\text{MLP}^i$, is rated by a differentiable loss function, which is the same as used in Ref.~\cite{jung2023_PRL}. It includes the mean absolute error, but also additional terms which penalize deviations between the predicted and the true variance, as well as spatial correlations of the propensities (see SM of Ref.~\cite{jung2023_PRL}). We use $N_S=300$ initial structures, which are equally divided into a training and a test set. For the training we apply stochastic gradient descent with an Adam optimizer~\cite{Adam}. 

The training of tGlassMLP is performed in two steps. First, we train in an `equal-time' mode, meaning that separate networks are trained for given values of $\langle \bar{\mathcal{C}}_B(t) \rangle$ (i.e., at equal times relative to the structural relaxation time), but different temperatures $T \geq T_\text{min}$. We found that these individual networks transfer better to lower temperatures than the ones who were directly trained on all state points.

These individual networks are then applied to lower temperatures $ T < T_\text{min}$ and the average of the predicted propensity, $\overline{\mathcal{X}_\text{MLP}} = \frac{1}{N_1}\sum_i^{N_1} \mathcal{X}_\text{MLP}^i,$ is calculated. We have found that depending on the initial condition of the training, not all resulting networks are equally efficient. We keep all networks which fulfill $ |\overline{\mathcal{X}_\text{MLP}} - \langle \bar{\mathcal{C}}_B(t) \rangle| < 10^{-4}$ for further processing. This is a self-consistency test, as the average predicted propensity is an input data. The networks which satisfy the above criterion are used to predict propensities for the low temperature configurations $ T < T_\text{min}$ at the values of $\langle \bar{\mathcal{C}}_B(t) \rangle$ for which MD results are not available (typically low values of $\langle \bar{\mathcal{C}}_B(t) \rangle$). Similarly to knowledge distillation in machine learning, we include them into the training procedure of the full and final tGlassMLP model. See Table \ref{tab:overview} in the App.~\ref{ap:training} for a more detailed description. 

In the second training step, a single tGlassMLP network is trained using data from all times and temperatures, including the extrapolated data at $ T < T_\text{min}$ produced in the first step. As before,  we only retain networks such that $ |\overline{\mathcal{X}_\text{MLP}} - \langle \bar{\mathcal{C}}_B(t) \rangle| < 10^{-4}$ {for predicted propensities at $T < T_\text{min}$.} Out of 16 networks trained initially in this way in the second step, four networks were selected for the predictions shown in the remainder of this manuscript. Error bars on dynamic quantities calculated from the predictions of tGlassMLP correspond to the variance between the predictions made by these individual networks. 

Additional details on hyperparameters, values for $\langle \bar{\mathcal{C}}_B(t) \rangle$ and temperatures used for training are presented in \cref{ap:training}. The training of tGlassMLP was performed on a Laptop GPU (NVIDIA T600 Laptop) and with a total computational cost of about one day. This includes the training of all 16 individual networks per time scale in the first training step, and of 16 networks using the full data for all times and temperatures in the second step.

In the following we present our results, which are organised in four parts. In the first part we validate the performance and the transferability of tGlassMLP using results known from MD simulations. In the second part, we use the trained models to predict structural relaxation at the experimental glass transition temperature which is not accessible by computer simulations. The third and fourth parts include a detailed analysis of dynamic correlation lengths and of the properties of the trained tGlassMLP network itself.

\section{Validation of tGlassMLP}

\label{sec:performance}

\begin{figure}
	\includegraphics[width=\linewidth]{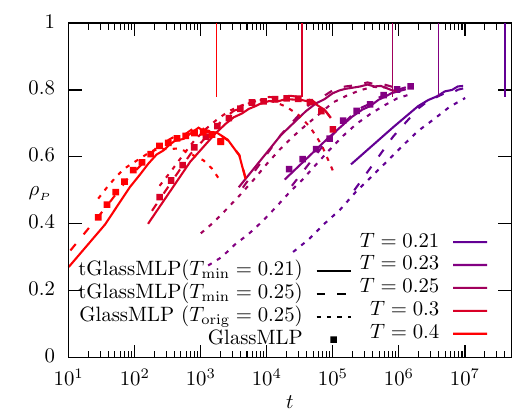}
	\caption{Pearson correlation coefficient of several trained ML models with MD simulations. Squares correspond to results shown in Ref.~\cite{jung2023_PRL} for GlassMLP trained at each state point. Dotted lines correspond to a GlassMLP model trained at $t=\tau_\alpha$ and $T_\text{orig} = 0.25$ and transferred to different times and temperatures. Full and dashed lines describe results for two transferable tGlassMLPs trained until minimum temperatures $T_\text{min} =0.21$ and $T_\text{min} =0.25$. Vertical lines indicate the structural relaxation times $\tau_\alpha^\text{BB}$ at each temperature.}
	\label{fig:pearson}
\end{figure}

We follow common practice~\cite{bapst2020unveiling} and investigate the performance of the trained model using the Pearson correlation coefficient,
\begin{equation}
\rho_P = \text{cov}(\mathcal{C}^i_B, \mathcal{X}^i_\text{MLP}) / \sqrt{\text{var}(\mathcal{C}^i_B) \text{var} (\mathcal{X}^i_\text{MLP})},
\end{equation}
which quantifies the correlation between the predicted propensities and the ground truth extracted from the MD simulations. The Pearson correlation varies between $\rho_P=1$ (perfect correlation) and $\rho_P=-1$ (perfect anti-correlation), while no correlation yields $\rho_P=0$.  

We find that the maximum of the Pearson correlation reaches values of roughly $\rho_P^\text{max} \approx 0.8$ for the lowest temperatures, thus indicating very strong correlations (see Fig.~\ref{fig:pearson}). Most importantly, we observe that the performance of the trained tGlassMLP networks (full lines) is as good as the results reported in Ref.~\cite{jung2023_PRL} in which the networks were individually trained on each state point (squares). This is a significant result considering that the tGlassMLP network only uses twice the number of fitting parameters to describe more than 50 different state points. This shows that the transferable network indeed discovers common features in the description of structural relaxation from the microscopic structure at different times and temperatures.

We have also trained tGlassMLP using a smaller dataset down to $T_\text{min}=0.25$ (dashed line in Fig.~\ref{fig:pearson}). The model very favorably transfers to lower temperatures and can predict structural relaxation on time scales that are orders of magnitude longer than in the training dataset with nearly as much accuracy as the directly trained model. This demonstrates excellent transferability of tGlassMLP in the regime where this can be tested quantitatively.

\begin{figure}
	\includegraphics[width=\linewidth]{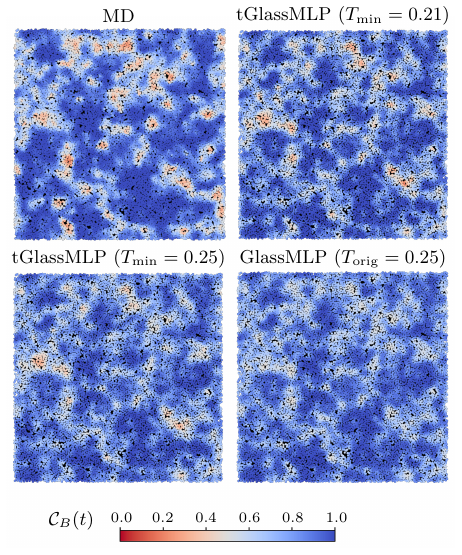}
	\caption{Snapshots comparing the spatial variations of the structural relaxation at $T=0.21$ and $\langle \bar{\mathcal{C}}_B \rangle=0.8$ obtained in MD simulations, to three different models: tGlassMLP networks trained using $T_{\rm min}=0.21$ and $T_{\rm min}=0.25$, and GlassMLP trained at $T_{\rm orig}=0.25$ and $t=\tau_\alpha^\text{BB}$. 
 }
	\label{fig:snapshots_T021}
\end{figure}

In addition, tGlassMLP outperforms the directly transferred GlassMLP model which was trained at the state point $t=\tau_\alpha$ and $T_\text{orig} = 0.25$ (dotted lines). This conclusion similarly holds when comparing the results to other ML techniques such as GNNs~\cite{bapst2020unveiling,pezzicoli2022se}. In particular, the equivariant network proposed in Ref.~\cite{pezzicoli2022se} shows very comparable transferability as the original GlassMLP network~\cite{jung2023_PRL}, but cannot match the improved transferability performance of tGlassMLP.

To visualize the correlation between the propensities obtained from MD simulations and the different ML models we show snapshots for a large configuration with $N=25800$ at $T=0.21$ in \cref{fig:snapshots_T021}. The tGlassMLP models are very accurate in predicting the location of both strongly rearranging regions and frozen regions in which no rearrangements take place. Stronger differences with the MD result are observed when using the GlassMLP network trained at $T_\text{orig} = 0.25$ and transferred to $T=0.21$. While also for this model, the propensities are correlated with the MD results, dynamic heterogeneities are less pronounced and the contrast between active and frozen regions is not captured properly. 

\begin{figure}
	\includegraphics[width=\linewidth]{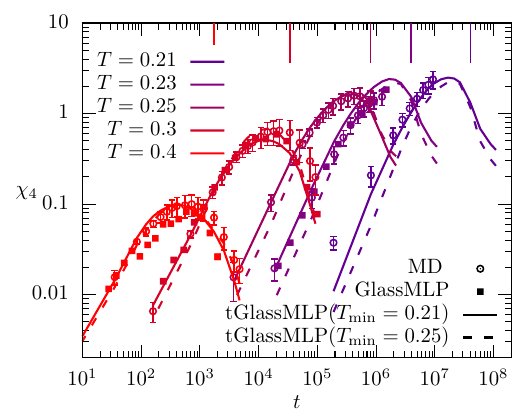}
	\caption{Dynamic four-point susceptibility $\chi_4(t)$. Points are MD results, squares are results shown in Ref.~\cite{jung2023_PRL} for GlassMLP trained at each state point. Full and dashed lines correspond to transferable tGlassMLPs trained until minimum temperatures $T_\text{min}=0.21$ and $T_\text{min}=0.25$. Vertical lines indicate $\tau_\alpha^\text{BB}$ at each temperature.}
	\label{fig:susceptibility}
\end{figure}

To analyze spatial correlations of the propensity quantitatively, we compute the four-point susceptibility $\chi_4(t)$,
\begin{equation}
  \label{eq:susceptibility}
\chi_4(t)= N_1 \left( \langle \bar{\mathcal{C}}_B^2(t)  \rangle - \langle \bar{\mathcal{C}}_B(t)  \rangle^2  \right),
\end{equation}
which was used extensively to characterize dynamic heterogeneity in supercooled liquids~\cite{bookDH}. As expected after inspection of the snapshots in \cref{fig:snapshots_T021}, we find very good agreement between the susceptibilities extracted from the MD results and the trained ML model (see Fig.~\ref{fig:susceptibility}). The results are as good as those reported in Ref.~\cite{jung2023_PRL} for GlassMLP models trained at each state point. In fact, for $T=0.4$, the transferable tGlassMLP model even surpasses the performance of GlassMLP, despite being much more broadly applicable to dynamic susceptibilities which can differ by more than two orders of magnitude in amplitude. 

This conclusion holds for the two models trained using different values of $T_\text{min}$. Therefore, the tGlassMLP model is able to predict a realistic increase in dynamic heterogeneity with decreasing the temperature even when it is extrapolated beyond the range used in the training set. This excellent transferability sets it apart from techniques which are trained on a single state point and are therefore unable to capture variations in the overall dynamic heterogeneity. The tGlassMLP models also predict a realistic decay of the susceptibility for times longer than the structural relaxation time even though for $T < 0.25$ no training data for times $t > 0.3 \tau_\alpha^\text{BB}$ were used during training. The models are therefore not only transferable in temperature, but also in time.

The above analysis clearly highlights that the tGlassMLP network is able to extrapolate its predictions outside of the range covered in the training set. It is therefore tempting to train tGlassMLP with as much data as is currently possible using MD simulations, and then use the network to predict features of structural relaxation at times and temperatures where MD simulations can no longer be performed. This is investigated in the next section. 

\section{Predicting the dynamics at the experimental glass transition temperature}

\label{sec:glass_transition}

The glass transition temperature $T_g$ is conventionally defined as the temperature at which structural relaxation occurs roughly $10^{12}$ times slower than in the simple liquid. Simulating the relaxation of supercooled liquids at $T_g$ would require integrating about $10^{14}$ time steps. The computational cost of the data provided in this section would thus roughly amount to $10^{11}$\,CPU\,hours. Equivalently, this would require about 100 days of simulation by completely exhausting the Top 500 supercomputers in the world.

Our approach is obviously more parsimonious and combines two algorithms: (i) the swap Monte Carlo algorithm allows us to efficiently create independent equilibrium configurations of our glass-model down to $T_g$, and (ii) the transferable tGlassMLP model which can predict dynamic propensities for each particle from the equilibrium configurations obtained in (i). This unique combination enables us to predict and analyze the strength, geometry and length scale of dynamic heterogeneity at unprecedentedly low temperatures and large times. 

\begin{figure}
	\includegraphics[width=\linewidth]{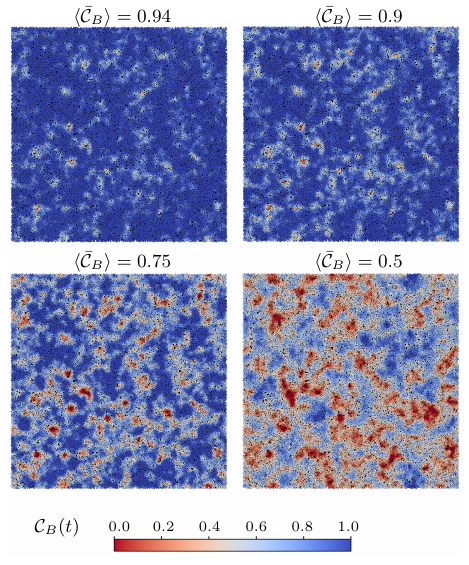}
	\caption{Snapshots visualizing the time evolution of the structural relaxation at $T_g=0.15$, with the corresponding value of the average $\langle \bar{\mathcal{C}}_\text{B} \rangle$ indicated, corresponding to estimated times $t=10^{11}, 3 \times 10^{11}, 10^{12}, 3 \times 10^{12}$. The corresponding movie can be found in the SM~\cite{SM}.}
	\label{fig:snapshots_transfer}
\end{figure}

In Fig.~\ref{fig:snapshots_transfer} we show snapshots describing the predicted time evolution of the structural relaxation in a given sample at $T=T_g$ with $N=82560$ using the tGlassMPL model. The same process is displayed as a movie in the SM, which better highlights how relaxation sets in at specific localised regions within the amorphous structure before slowly spreading over the entire system. This qualitative description is similar to results produced by MD simulations performed for a lesser degree of supercooling.

Since these predictions are made for temperatures where direct MD simulations can no longer be performed, there is no way to directly test the quality of these predictions. The plausibility of the result is guaranteed by the excellent transferability demonstrated in the previous section for a higher temperature regime and the physically consistent behaviour that is predicted. 

\begin{figure}
  \includegraphics[width=\linewidth]{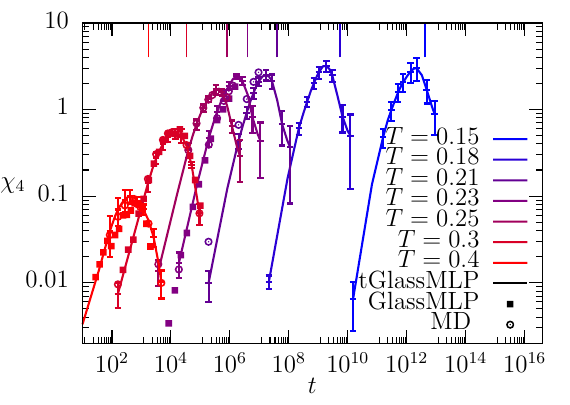}
	\caption{Dynamic four-point susceptibility $\chi_4(t)$ for tGlassMLP (using $T_{\rm min}=0.21$). Square data points correspond to results shown in Ref.~\cite{jung2023_PRL} and points to MD results. Vertical lines indicate $\tau_\alpha^\text{BB}$ at each temperature. Error bars correspond to the variance between individual networks, trained using the same dataset but with different initial weights.
 }
	\label{fig:susceptibility_transfer}
\end{figure}

We quantify the dynamic heterogeneities visible in these snapshots using the four-point susceptibility $\chi_4(t)$, defined in \cref{eq:susceptibility}. The results are presented in Fig.~\ref{fig:susceptibility_transfer}. The time-dependence is extracted by predicting the value of $C_B^i$ for each particle for a given value of the average quantity $\langle \bar{\mathcal{C}}_\text{B} \rangle$, and converting $\langle \bar{\mathcal{C}}_\text{B} \rangle$ into a time using time-temperature superposition, as discussed above in Sec.~\ref{sec:observables}. The predicted $\chi_4(t)$ functions continue to have the non-monotonic time dependence they have at higher temperatures, and we note that the amplitude of the maximum of $\chi_4$ grows very modestly for temperatures $T < 0.21$ (see \cref{fig:susceptibility_transfer}). It should be emphasized that this is not a trivial result. Indeed in \cref{sec:performance} we have observed that tGlassMLP with $T_\text{min}=0.25$ has correctly predicted a slowly-increasing $\chi_4$ when transfered beyond the temperature range it has been trained on.

\section{Dynamic length scales}

\label{sec:length_scales}

\subsection{Four-point dynamic length scale}

The dynamic heterogeneities visualized in \cref{fig:snapshots_transfer} and quantified via $\chi_4(t)$ in \cref{fig:susceptibility_transfer} can also be characterized by a dynamic correlation length, which describes spatial correlations of the dynamic propensity. Calculating static and dynamic length scales, in particular at very low temperatures, is of high importance in glass physics, since the emergence and growth of dynamic correlation length scales is a key element of various different theories of the glass transition~\cite{bennemann1999growing,Flenner2010,RevModPhys.83.587,Berthier2022}.

Here, we first adopt the same methodology as used in Ref.~\cite{jung2023_PRL} to extract a dynamic length scale from the four-point dynamic structure factor~\cite{Flenner2010},
\begin{align}
S_4(q,t) &= N_1^{-1} \left\langle W(\bm{q},t) W(-\bm{q},t) \right\rangle\\
 W(\bm{q},t) &= \sum_{i \in N_1}  (\mathcal{C}_B^i(t) - \langle \bar{\mathcal{C}}_B(t) \rangle) \exp[\textrm{i} \bm{q}\cdot \bm{R}_i(0)]. \nonumber
\end{align}
The dynamic structure factor $S_4(q,t)$ characterizes the geometry and spatial extent of the regions defined by correlated fluctuations of $\mathcal{C}_B^i(t)$. In the limit $q \rightarrow 0$, it has been predicted to decay quadratically with $1/q$, according to the Ornstein-Zernicke form \cite{Flenner2010},
\begin{equation}
S_4(q,t) \approx \frac{\Tilde{\chi}_4}{1 + (q\xi_{4}(t))^2}.
\end{equation}
Here, $\Tilde{\chi}_4$ is connected (but not equal) to the dynamical susceptibility $\chi_4$~\cite{Flenner2010} defined above, and $\xi_{4}(t)$ denotes the time-dependent dynamic length scale. This correlation length can be extracted by fitting the measured dynamic structure factor for small $q$. Recently, we have applied this method to GlassMLP networks by utilizing the scalability of the network in system size~\cite{jung2023_PRL}. This last step was necessary, since an accurate extraction of $\xi_{4}$ requires very large system sizes with linear size $L_S \gg \xi_4$~\cite{karmakar2010analysis,Flenner2010,karmakar2014growing,Flenner2016}. Here, we apply the same methodology as in Ref.~\cite{jung2023_PRL}, using the same fitting parameters, to extract the dynamical length scale $\xi_{4}$ from systems with $N=82560$ particles, but we now use tGlassMLP models.

\begin{figure}
	\includegraphics[width=\linewidth]{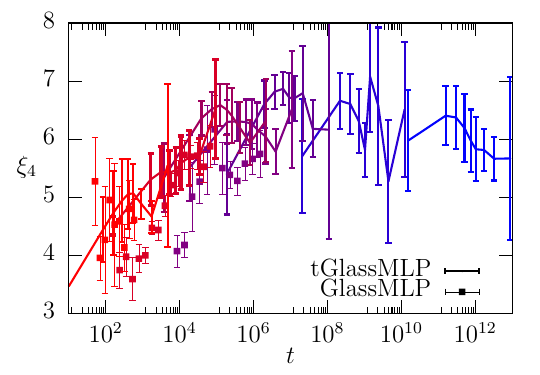}
	\caption{Four-point dynamic length scale $\xi_{4}(t)$ extracted from $S_4(q,t)$. Results are shown for the tGlassMLP model with $T_{\rm min}=0.21$. We also show the length scale extracted in Ref.~\cite{jung2023_PRL} using GlassMLP. Colors are the same as in \cref{fig:susceptibility_transfer}.}
	\label{fig:xi}
\end{figure}

The results for $ \xi_{4} $ are shown in \cref{fig:xi}. At any given temperature, the length scale increases with time up to a maximum value which is located roughly at the structural relaxation time, $\tau_\alpha^\text{BB}$. This maximum increases steeply with decreasing the temperature up to $T=0.25$, below the growth is much less pronounced to reach maximal values of the order $\xi_{4} \approx 6-7$ at most. This result is consistent with the findings in Ref.~\cite{jung2023_PRL} for $T \leq 0.23$, and is also consistent with the slow increase of the dynamic susceptibility $\chi_4$ in \cref{fig:susceptibility_transfer}.

Similarly to Ref.~\cite{jung2023_PRL} we have also extracted a higher-order contribution to the dynamic structure factor, as shown in App.~\ref{ap:A}. Its evolution with time and temperature again mirrors the behaviour of $\chi_4$ and $\xi_4$. 

\subsection{Average chord length}

To enable a more detailed comparison of these results to MD simulations we propose an alternative definition for a dynamic length scale, which is connected to the average chord length recently defined in Ref.~\cite{Berthier2022}. One advantage of the chord length is that is does not require very large system sizes, in contrast to the extraction of $\xi_4$ which does~\cite{karmakar2010analysis,Flenner2010,karmakar2014growing,Flenner2016}. To calculate the average chord length, particles are mapped onto a discrete lattice, and the propensity $a$ of each lattice site is defined as the average propensity of the particles within the lattice site. If the propensity of a site is $a < 0.5$, it is denoted as mobile, and immobile otherwise. After discretisation and thresholding, chords are defined as series of adjacent mobile sites along all rows and columns of the lattice. The linear size of a chord is $\xi_\text{chord}$, and the average over all chords provides the average chord length $\langle \xi_\text{chord} \rangle$. See Section VI. D and Fig.~18 in Ref.~\cite{Berthier2022} for further explanations and visualization. 

We adapt the definition of the lattice propensity $a$ to make the resulting chord length more comparable to $\xi_{4}$, in particular for times of the order of $\tau_\alpha^\text{BB}$. {For each time and temperature, we calculate the median of the bond-breaking propensities for all particles. If a particle propensity is below this median it is defined as active ($a^i=0$), and passive otherwise ($a^i=1$). 
Afterwards, we follow the same procedure as described above by mapping the newly defined dynamical quantity, $a^i,$ onto a lattice, thus redefining the lattice propensity $a$}, and calculating the chord length $\xi_\text{chord}$. This leads to a different time evolution compared to the definition in Ref.~\cite{Berthier2022}, which features a monotonic growth of the average chord length with time as relaxed regions gradually fill the entire system. In the present version, only the clustering of the mobile particles contributes to the calculation of the chord length. 

\begin{figure}
	\includegraphics[width=\linewidth]{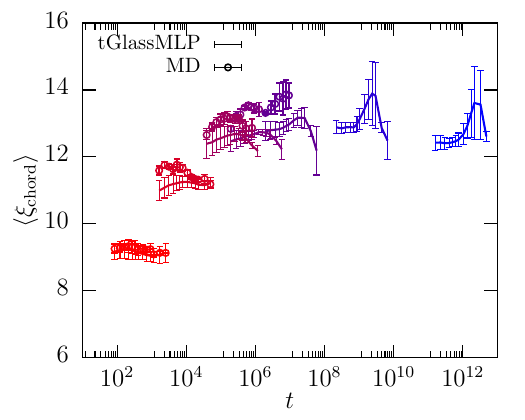}
	\caption{Average chord length, $\langle \xi_{\rm chord} \rangle$, {which can be used to measure dynamic length scales.} Colors are the same as in \cref{fig:susceptibility_transfer}. For each temperature we show times in the range $0.95 > \langle \bar{\mathcal{C}}_B(t) \rangle > 0.3.$}
	\label{fig:chord_length_50}
\end{figure}

We show the results for the time and temperature evolution of $\langle \xi_{\rm chord} \rangle$ in \cref{fig:chord_length_50}. As anticipated, the average chord length behaves similarly to $\xi_4$ (compare with \cref{fig:xi}). {Both quantities display a non-monotonic dependence in time, although the non-monotonicity is less pronounced for    $\langle \xi_{\rm chord} \rangle$. Most importantly, both length scales display the same temperature dependence when evaluated in the order of the structural relaxation time $\tau_\alpha^\text{BB}$. This result therefore establishes $\langle \xi_{\rm chord} \rangle$ as an interesting measure to extract dynamic length scales, also from smaller systems and for times $t<\tau_\alpha^\text{BB}.$ 
Consequently, we can also calculate $\langle \xi_{\rm chord} \rangle$ from the MD simulations performed to train tGlassMLP.} The MD results are in good agreement with the predictions of the ML model (see \cref{fig:chord_length_50}). The MD length scales are systematically slightly larger than the ones extracted using tGlassMLP. This offset, however, does not grow systematically in temperature. The MD results therefore draw the same picture of a dynamic correlation length which does not grow significantly beyond $ \langle \xi_{\rm chord} \rangle > 14$.

\section{Extracting physical information from the network}

\label{sec:network}
A natural question when applying machine learning to science is whether one can interpret what the neural network learns in order to achieve its task, and whether it is possible to extract some physical information from the trained network. In the following we will focus on this question for tGlassMLP. 

\subsection{Attention layer}

We presented in \cref{sec:methodology} the structure of the tGlassMLP network, where we emphasized the addition of an attention layer, which enables tGlassMLP to adapt the weight of different structural descriptors depending on the temperature. We now study these weights explicitly, and relate the observed evolution to the dynamic length scales discussed in Sec.~\ref{sec:length_scales}.

\begin{figure}
	\includegraphics[width=\linewidth]{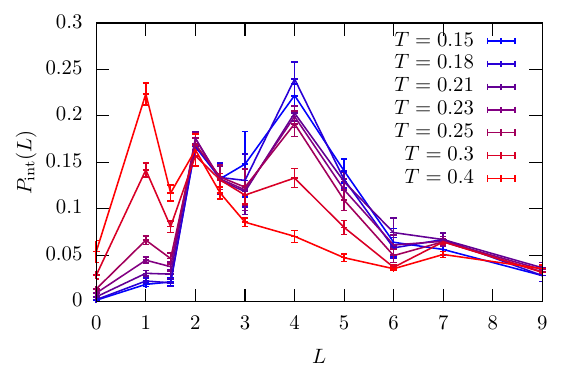}
	\caption{Normalised distribution of statistical weights $P_\text{int}(L)$, determined from the average weights of the attention layer of the tGlassMLP model with $T_\text{min}=0.21$ for each coarse-graining length $L$. Results are shown for different temperatures $T$ and a single time corresponding to $\langle \bar{\mathcal{C}}_B \rangle = 0.8$. {These distributions can be used to define an intrinsic, temperature-dependent dynamic length scale $\xi_\text{int}(T) = \int \text{d}L P_{int}(L) L$}.}
	\label{fig:intrinsic_length}
\end{figure}

After training, each of the physics-inspired structural input descriptors is assigned a statistical weight which depends on time and temperature. To relate these weights to a length scale, we first calculate the particle-averaged weight of each descriptor, and average this result over all descriptors which are coarse-grained over the same length scale $L$. The resulting normalized averaged weights for each $L$ value are then used to construct the function $P_\text{int}(L)$. This function is estimated after training the network, and depends on  temperature $T$.   

The temperature evolution of the distribution $P_{\rm int}(L)$ is shown in Fig.~\ref{fig:intrinsic_length}. One can clearly observe how the network puts increasing weight on larger length scales when the temperature is decreased. This evolution demonstrates that descriptors which are coarse-grained over larger length scales have an increased weight at lower temperatures. 
This implies that the network learns the existence of a length scale growing at low temperature and thus determining the range over which structural  heterogeneities matter for dynamical behaviors.

We have studied this length scale $\xi_\text{int}(T) = \int \text{d}L P_{int}(L) L$ (recall that $P_{\rm int}(L)$ is normalised). The temperature evolution of $\xi_\text{int}(T)$ compares well to the other dynamic length scales, as shown below in Fig.~\ref{fig:length_scale} and discussed further in Sec.~\ref{sec:discussions}. This result confirms, \emph{a posteriori}, the utility of the attention layer in tGlassMLP to represent different temperatures with different dynamical correlation lengths. Additionally, it gives an interesting interpretation from the perspective of glass physics to the weights learned by the model.

\subsection{Bottleneck layer}

\begin{figure}
\includegraphics[width=\linewidth]{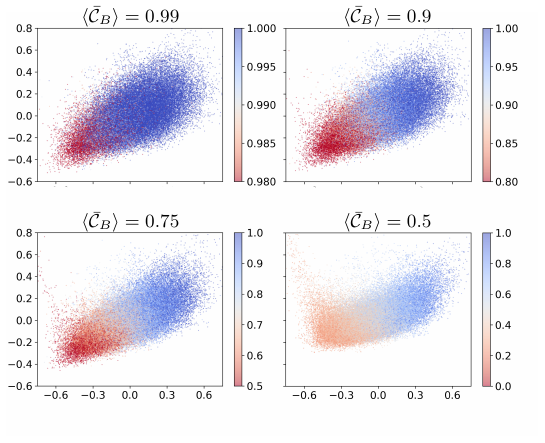}
	\caption{Visualization of the amorphous structure, encoded by the two-dimensional bottleneck layer, at $T=0.25$ for different times. Each point represents a single particle ($N=90000$ particles shown in total) at coordinates given by the value of the bottleneck layer. Each point is colored according to the value of the particle propensity $\mathcal{C}^i_B(t)$, as indicated by the color bar.}
	\label{fig:bottleneck_time}
\end{figure}

\begin{figure}[b]
	\includegraphics[width=\linewidth]{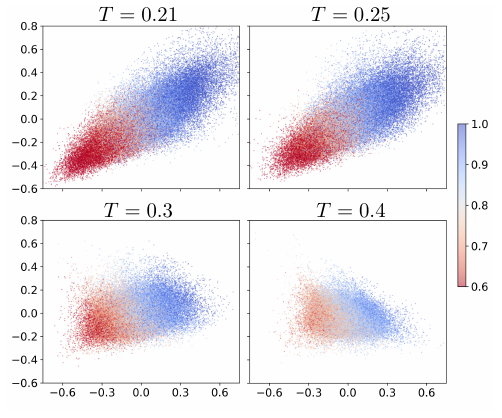}
	\caption{Visualization {of the amorphous structure, encoded by the} two-dimensional bottleneck layer at fixed time, $\langle\bar{\mathcal{C}}_\text{B} \rangle = 0.8$, and different temperatures. The figure is constructed as in \cref{fig:bottleneck_time}.}
	\label{fig:bottleneck_temp}
\end{figure}

Following the attention layer, the structural input is encoded into a two-dimensional bottleneck layer. This dimensional reduction implies that 
the dynamics mainly depends on two variables. However, those two variables are obtained by the non-trivial aggregation of several physical local features. Hence, it is unclear how to provide a simple interpretation of those two variables. It is even possible that such an interpretation does not exist. 

Still, we can provide some direct evidence of the relationship between these two variables and the dynamical behavior. The output of the bottleneck layer for $T=0.25$ and different time scales $\langle \bar{\mathcal{C}}_B(t) \rangle $ is shown in \cref{fig:bottleneck_time}. Each point in the figure represents a single particle, its coordinates correspond to the values of the bottleneck layer while the color is used to code for the value of the particle propensity extracted from MD simulations. {In this representation, we can observe a clear separation of structure into active and passive particles. All the intricacies of amorphous structure are therefore hidden in the complex encoding from the structural descriptors to the bottleneck layer. Interestingly, in this description structural relaxation over time is `simple'. Indeed, the figure clearly demonstrates how the red color, corresponding to the most mobile particles, gradually spreads from the bottom left corner towards the top right. Thus, at early times, only those particles with small values in both bottleneck nodes are likely to rearrange. At longer times, more particles become mobile even at larger values of the bottleneck nodes, until eventually all particles rearrange at very long times. 

We also study the temperature dependence of the bottleneck layer for a fixed  timescale relative to $\tau_\alpha^\text{BB}$ in \cref{fig:bottleneck_temp}. The aggregate output of the bottleneck layer is slightly deformed by the change in temperature, which arises from the temperature-dependent weights in the attention layer. More importantly, however, we observe that the separation between active and passive particles becomes more pronounced at lower temperatures, which indicates some sort of increasing structural heterogeneity. This observation explains the improved performance of tGlassMLP at lower temperatures, as quantified by the Pearson correlation in \cref{fig:pearson}. 

\section{Discussion and conclusion}

\label{sec:discussions}

We presented and analyzed the tGlassMLP model, which can predict the relaxation dynamics of deeply supercooled liquids from the amorphous microstructure over a large range of time scales and temperatures. The approach has been verified by calculating Pearson correlation coefficients $\rho_P$ with direct MD simulations and by evaluating dynamic susceptibilities $\chi_4$ and dynamic correlation length scales $\xi_4$. 

Overall, the predictions of tGlassMLP regarding spatial correlations support a scenario in which dynamic heterogeneities do not grow significantly when the temperature is decreased far below $T_\text{MCT}$ and approaches the experimental glass transition $T_g.$ To summarize these results, we compare in Fig.~\ref{fig:length_scale} the different techniques used above to calculate dynamic length scales at a time $t$ corresponding to $\langle\bar{\mathcal{C}}_{B}(t) \rangle = 0.8$. We scale the results with a method-dependent factor and thus find very good overlap between all results~\footnote{The factor of $1.5$ to scale $\xi_\text{int}$ is determined empirically, and depends on the absolute weights used in the MLP to encode the output of the attention layer to the bottleneck layer. The weights of the attention layer can obviously only capture relative changes of the overall length scale with temperatures. Similarly, the factor of 2 for the chord length is reasonable since chords represent the linear extension of mobile regions, and $\xi_4$, approximately, their radius.}. The figure shows that all methods produce the same trend of a dynamic length scale increasing quite rapidly at high temperatures, followed by a much slower evolution when $T < T_\text{MCT} = 0.3$. The length scales directly determined by MD follow the same trend, although of course on a much smaller temperature range. We also found that the evolution of the dynamic susceptibility $\chi_4(t)$ follows a trend similar to the length scales shown in Fig.~\ref{fig:length_scale}.

\begin{figure}
	\includegraphics[width=\linewidth]{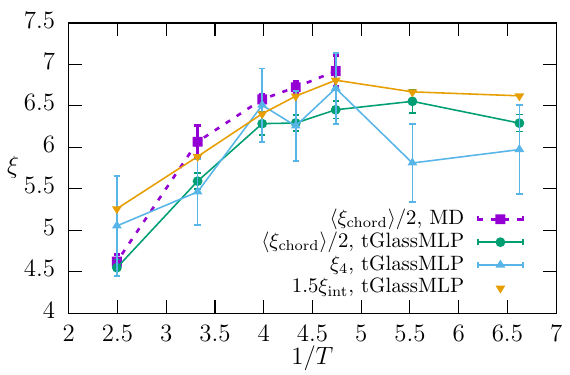}
	\caption{
 Temperature-dependent length scales $\xi$ as extracted from various techniques described in the manuscript at $\langle\bar{\mathcal{C}}_\text{B} \rangle = 0.8$. The figure demonstrates a good agreement between the different lengths. 
 }
	\label{fig:length_scale}
\end{figure}

Our data-driven approach to predict dynamic heterogeneity is quite general and is based on detecting structural heterogeneity which can easily be measured using the swap Monte Carlo algorithm at very low temperatures. Nevertheless, it still leaves open the possibility that the extrapolated tGlassMLP networks do not capture all important mechanisms leading to growing heterogeneities at low temperatures. However, taking the extrapolated data at face value, it is interesting to compare the emerging scenario in which dynamic heterogeneities only grow very weakly below $T<T_\text{MCT}$ to available numerical and experimental data. 

Several articles have featured strongly increasing dynamic heterogeneities and correlation lengths, including colloidal experiments \cite{weeks2000three} and computer simulations~\cite{bennemann1999growing,berthier2004dynamic,charbonneau2013decorrelation,Flenner2010} in the high temperature regime. This is in agreement with our results. The change of this behavior toward a weaker increase at temperatures lower than $T_\text{MCT}$ is also found in several simulation and experimental studies. In the three-dimensional Kob-Andersen mixture a crossover from strongly increasing dynamic heterogeneities to weakly-increasing has recently been observed~\cite{hocky2014crossovers,coslovich2018dynamic,DAS2022100098,10.1063/1.4773321,Berthier2007_1,Berthier2007_2}. Also in other models, such as binary mixtures of quasi-hard spheres~\cite{kob2012non} it has been observed that dynamic heterogeneities do not continue to grow significantly. A plateau and a crossover to a weakly increasing heterogeneity has been observed in Ref.~\cite{ortlieb2021relaxation} for Kob-Andersen binary mixtures over a range of mixing ratios. Only very recently were measurements performed at much lower temperatures in a two-dimensional soft sphere system~\cite{Berthier2022}, but here the saturation was less pronounced. These above numerical findings in three dimensions are consistent with experiments on molecular glass-formers near the experimental glass transition temperature, which show length scales in the order of 5 molecular diameters~\cite{ediger2000spatially,doi:10.1126/science.1120714} and weakly increasing dynamic heterogeneity~\cite{dalle2007spatial,albert2016fifth}. 

The analysis of the intrinsic properties of the trained tGlassMLP model shows that the effect of structure on dynamics can be reduced to just two variables. However,  we could not find any direct interpretation for them, because they are an aggregate of several physical local features. Despite their good predicting power, there is no clear separation between mobile and immobile particles in this two dimensional representation, thus casting doubts on the possibility of finding a connection between simple form of amorphous order or simple local structures and dynamics by unsupervised learning (similar `potato-shaped' dimensional reductions were found in various glass-formers in Ref.~\cite{doi:10.1063/5.0128265}.) It would be interesting to study whether this conclusion applies to the dynamics observed in individual trajectories~\cite{doi:10.1126/science.1166665,PhysRevLett.119.028004} instead of the dynamic propensity.

Finally, we believe that the transferable ML approach presented in this work can be an important step towards the study of dynamic heterogeneities in a manifold of glass-forming materials. This includes models in different spatial dimensions, continuously polydisperse models~\cite{Berthier2022} as well as more fragile glass-formers, in which dynamic heterogeneities might show a different fate at low temperatures~\cite{ROYALL201534}. Furthermore, quantifying the performance of transferability is an important topic in the machine learning community~\cite{bao2019transferability,tran2019transferability}. Future work could include advanced transferability measures~\cite{nguyen2020leep} or self-supervised learning~\cite{gidaris2018unsupervised}. Employing such techniques to our model would be an important step to strengthen further the reliability of transferred results.

\acknowledgments

We thank D. Coslovich and F. Landes for useful discussions. This work was supported by a grant from the Simons Foundation (\#454933, Ludovic Berthier, \#454935 Giulio Biroli).

\FloatBarrier

\appendix

\section{Conversion between times and bond-breaking correlation values}

\label{ap:BB}

In \cref{sec:observables} we introduced the bond-breaking correlation and explained how the average value $\langle \bar{\mathcal{C}}_B(t) \rangle$ is used to encode time in the machine learning procedure. Here, we provide further evidence to support this approach.

\begin{figure}
	\includegraphics[scale=1.1]{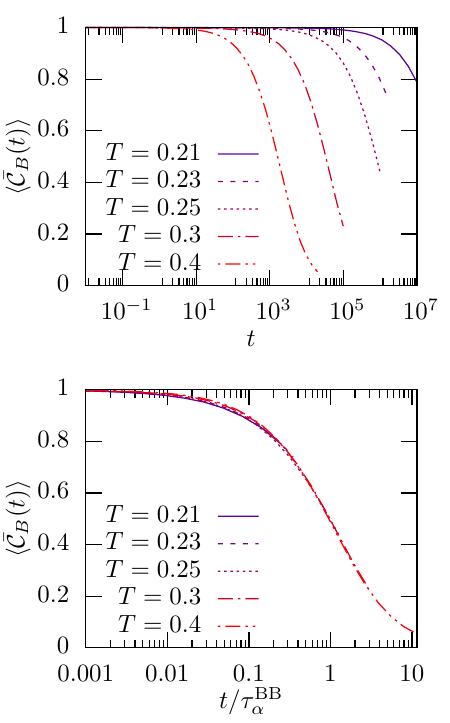}
	\caption{Average bond-breaking correlation $\langle \bar{\mathcal{C}}_B(t) \rangle$ for different temperatures. Top and bottom show the same results, where bottom is rescaled by the estimated structural relaxation time, $\tau_\alpha^\text{BB}$.}
	\label{fig:BB}
\end{figure}

In \cref{fig:BB} it can be observed that the decay of $\langle \bar{\mathcal{C}}_B(t) \rangle$ at various temperatures is very similar, although the curves at different temperatures decay on different time scales. We extract the structural relaxation time $\langle \bar{\mathcal{C}}_B(t=\tau_\alpha^\text{BB}) \rangle = 0.5$ also for the curves which are not simulated long enough by maximizing the overlap between the curves. The result is shown in bottom panel of \cref{fig:BB}. The curves display nice time-temperature superposition, indicating that the shape of the time decay changes very little (if at all) over a wide temperature range.  

Based on the extracted values for $\tau_\alpha^\text{BB}$, which show an Arrhenius dependence above $T \leq 0.25$, we extrapolate the corresponding values for $\tau_\alpha^\text{BB}$ at lower temperatures and find that the glass transition temperature $\tau_\alpha^\text{BB}(T=T_g) = 10^{12}$ can be estimated as $T_g=0.15,$ which is the lowest temperature considered in the manuscript. To reconstruct the time-dependence of propensity at such low temperatures, we predict propensity at a certain value of $\langle \bar{\mathcal{C}}_B \rangle$, use \cref{fig:BB} to estimate $t/\tau_\alpha^\text{BB}$ and thus deduce $t$ using the previously estimated value for $\tau_\alpha^\text{BB}$.

\section{Details on tGlassMLP and the training procedure}

\label{ap:training}

The core of tGlassMLP is the $M_S$-dimensional input layer, followed by the attention layer, as described in \cref{sec:methodology} and sketched in \cref{fig:geometry}. The state point is inserted into the attention layer via MLP1. MLP1 has a six-dimensional input layer, in which we feed the slightly transformed state point, ($1/T$, $1/T^2$, $1/T^3$, $\langle \bar{\mathcal{C}}_B(t) \rangle$, $\langle \bar{\mathcal{C}}_B(t) \rangle^2$, $\langle \bar{\mathcal{C}}_B(t) \rangle^3$), and two six-dimensional hidden layers, each with an ELU activation function~\cite{ELU}. The two-dimensional output is then calculated using a linear activation function.

After the attention layer, the data is encoded into the two-dimensional bottleneck layer, which is concatenated with the output by MLP2, and followed by two ten-dimensional hidden layers, each with an ELU activation function. MLP2 is similar to MLP1 just formed of two seven-dimensional hidden layers and a four-dimensional output layer. The performance of tGlassMLP, however, does not crucially depend on any of these choices. The final output $\mathcal{X}_\text{MLP}^i$ is then calculated using a linear activation function.

\begin{table}
	{
		\centering 
		\renewcommand{\arraystretch}{1.15}
		\textbf{tGlassMLP ($T_\text{min}$ = 0.21)}
		\begin{tabular}{|c|c|c|c|c|c|c|c|}
			\hline

			$ \langle\bar{\mathcal{C}}_B(t) \rangle$ $\diagdown$ $T$ & 0.15 & 0.18	& 0.21 & 0.23 & 0.25 & 0.3 & 0.4 \\
			\hline
			0.99 & t & t	& x & x& x  & x &  x\\
			\hline
			0.94 & t & t	& x & x& x  &  x& x \\
			\hline
			0.90 & t & t	& x & x& x  & x &  x\\
			\hline
			0.85 & t & t	& x & x& x  & x & x \\
			\hline
			0.8 & t & t 	& x & x  & x& x & x \\
			\hline
			0.75& r & t	& t & x& x  & x &x  \\
			\hline
			0.65& r &t	& t &t  & x & x & x \\
			\hline
			0.5	& r & t & t & t & x & x & x \\
			\hline
			0.25& r & r	&  r & r & r & x & x \\
			\hline
			0.15&&	&  &  &  &  & x \\
			\hline
			0.075&&	&  &  &  &  & x \\
			\hline
		\end{tabular}
	
	\vspace*{0.5cm}
	
	\textbf{tGlassMLP ($T_\text{min}$ = 0.25)}
	\begin{tabular}{|c|c|c|c|c|c|}
		\hline
		
		$ \langle\bar{\mathcal{C}}_B(t) \rangle$ $\diagdown$ $T$ 	& 0.21 & 0.23 & 0.25 & 0.3 & 0.4 \\
		\hline
		0.99  	& t & t& x  & x &  x\\
		\hline
		0.94  	& t & t& x  &  x& x \\
		\hline
		0.90  	& t & t& x  & x &  x\\
		\hline
		0.85  	& t & t& x  & x & x \\
		\hline
		0.8  	& t & t  & x& x & x \\
		\hline
		0.75 	& t & t & x  & x &x  \\
		\hline
		0.65 	& t &t  & x & x & x \\
		\hline
		0.5	  & t & t & x & x & x \\
		\hline
		0.25 	&  r & r & r & x & x \\
		\hline
		0.15	&  &  &  &  & x \\
		\hline
		0.075	&  &  &  &  & x \\
		\hline
	\end{tabular}
	}
\caption{Overview over the different state points used for the training of tGlassMLP. `x' denotes training data extracted from MD simulations, `t' are transfered results using the first training step in ``equi-time'' mode. `r' are results shown for the final trained tGlassMLP networks.}
\label{tab:overview}
\end{table}

As described in the main text, tGlassMLP is then trained in two steps: The first ``equi-time'' step for constant $\langle\bar{\mathcal{C}}_B(t) \rangle $ is followed by a self-consistency check and transfer procedure. The results are then used for the final training of the full tGlassMLP model. The data that is used for the tGlassMLP models presented in this work is summarized in \cref{tab:overview}.

\begin{table}
	\centering 
	\renewcommand{\arraystretch}{1.15}
	\begin{tabular}{|c|c|c|c|}
		\hline
		Phase 1	& 10e, $10^{-3}$ & 50e, $5\cdot10^{-4}$ & 25e, $2\cdot 10^{-4}$		 \\
		\hline
		Phase 2	&50e, $4\cdot 10^{-5}$  & \multicolumn{2}{c|}{} \\
		\hline
	\end{tabular}
	
	\caption{Accuracies of the Adam optimizer used for training. ``e'' stands for the number of epochs.}
	\label{tab:adam}
\end{table}

Each training step is performed in a similar way as the training of GlassMLP \cite{jung2023_PRL}.  The batch size is equal the number of type 1 particles per configuration, $N_\text{batch} = N_1 = 600.$ The training of both the ``equi-time'' step and the final step is separated into two phases. In the first phase, the model is trained for 85 epochs with a loss function that only considers the mean squared error. The accuracy of the Adam optimizer is varied as described in \cref{tab:adam}. Afterwards, the model is trained for another 50 epochs using the loss function with the same parameters as for GlassMLP.

When applying tGlassMLP on a large set of amorphous structures, in particular at low temperatures and long times, we use an iterative procedure to constrain the predicted mean propensity $ \overline{\mathcal{X}_\text{MLP}} = \frac{1}{N_1}\sum_{i=1}^{N_1} \mathcal{X}_\text{MLP}^i$ to the expected value $\langle \bar{\mathcal{C}}_B(t) \rangle.$ To achieve this we define $\mathcal{C}_B^0=\langle \bar{\mathcal{C}}_B(t) \rangle$ as the initial input in tGlassMLP. Using the output in iteration $j$, we update the input via the recursive relation $\mathcal{C}_B^{j+1}=\mathcal{C}_B^{j} + 0.75\cdot(\overline{\mathcal{X}_\text{MLP}}^j - \langle \bar{\mathcal{C}}_B(t) \rangle).$ We iterate for a maximum of 6 iterations, and check for convergence. The final $\mathcal{C}_B^{j}$ is never very different from $\langle \bar{\mathcal{C}}_B(t) \rangle$ due to the self-consistency check described in the main text.

\section{Higher-order term in four-point structure factor}

\label{ap:A}

\begin{figure}
	\includegraphics[width=\linewidth]{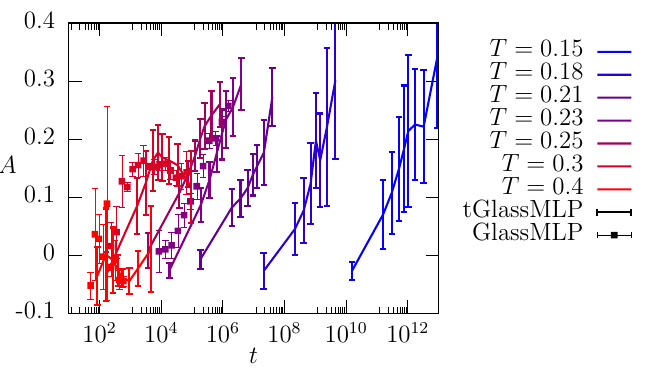}
	\caption{ Higher-order prefactor $A$, extracted from $S_4$ after fitting the length scale $\xi_{S_4}.$}
	\label{fig:A}
\end{figure}

In this appendix we analyze the higher-order prefactor $A$ as extracted from fitting the dynamic structure factor $S_4$ with an additional contribution which is of third order in $q$,
\begin{equation}
S_4(0.2<q<0.6,t) = \Tilde{\chi}_4(t) / \left( 1 +  (\xi q)^2 + A(\xi q)^3\right ).
\end{equation}
As discussed in Ref.~\cite{jung2023_PRL} this quantity shows a very pronounced increase around the mode-coupling temperature $T_\text{MCT}$, which we can perfectly reproduce using tGlassMLP (see Fig.~\ref{fig:A}). Now, we can also systematically study temperatures below $T_\text{MCT}$. We find that when entering deep into the glassy regime, also the prefactor $A$ reaches a plateau, indicating that the geometry of rearranging clusters is not significantly changing anymore, consistent with Ref.~\cite{DAS2022100098}.

\bibliography{library_local.bib}

\end{document}